\documentclass [fleqn,twoside,12pt]{article}
\usepackage{amsmath}
\usepackage{amssymb}

\begin{document}

\twocolumn[
\title{Model of embedded spaces: the field equations}
\author{Vitaly I. Noskov}
\date{\small \em Institute of Continuous
Media Mechanics, Ural Branch of Russian Acad. Sci., 1 Korolyov St., Perm 614013, Russia,
Email: nskv@icmm.ru}
\maketitle

\abstract
{A study of the Model of Embedded Spaces (MES) with a relativistic version
of Finslerian geometry is continued. The field equations of the MES
(Einstein and Maxwell types) are derived, and this formally completes
geometrization of classical electrodynamics. The minimal action principle
leads to geometrization of the field sources (the right-hand sides of the
equations) and, as a consequence, to a field hypothesis of matter, a
direct confirmation of W. Clifford's ideas.  }
\bigskip
\bigskip
]

\section{Introduction}

Judging by the state of the art, the idea of treating physical phenomena as
geometric ones in the really existing space-time of the Universe will be
long attractive for researchers. It was put forward by W. Clifford at the
end of the 19th century \cite{Kl} and partly proved by H. Minkowski
\cite{Mi} and A. Einstein in collaboration with M. Grossmann \cite{Ei1} in
the first quarter of the last century. The results of these works
essentially changed the general notions on the substance of entire areas of
physics. Geometrically, the essence of these works reduced to consecutively
changing the space-time models of physics. Obviously, these were the
first but uniquely effective steps toward geometrization of physics.

In our opinion, subsequent effort in this area did not lead to results
comparable with those cited above. The stumbling block was the problem of
geometrization of classical electrodynamics\footnote{This problem can be used
as a test problem for any attempts of geometrization of physics.}.

The main geometric idea behind the attempts of finding a solution to this
problem (as well as the attempts of geometrization of weak and strong
interactions) was to apply different types of more or less physically
grounded generalizations of the Riemannian geometry, based, e.g., on the
assumptions of sectional curvature \cite{Weil}, higher dimensions of real
space-time \cite{Kal,Kle,Ei2}, torsion \cite{Rod} and so on.  (Some new
results were obtained in the framework of the relational approach --- the
works on the theory of physical structures \cite{Kul,Vla}. Based on
their originally accepted principles, the authors deduced such realistic
objects as metric space-time and physically acceptable generalizations of
the Riemannian geometry, including versions with higher dimensions etc.)

To do justice to adherents of the ``Riemannian'' approach, we, however,
should give some arguments. In our opinion, the most productive is only
such a Riemannian generalization that is verified by experiment, though
partly, on {\it all\/} scales (macroscopic, molar and microscopic) because
the electromagnetic interaction is long-range. Besides, it is well known
that the possibly existing higher dimensions have a Planck scale and
must be compactified, which clearly contradicts the obvious first
judgement\footnote{We do not deny the opportunity of higher dimensions in
Riemannian or some other models, we only doubt whether geometrization of
physical phenomena of any scales can be realized using a geometry of the
Planck scale.}.

Thus we may suggest that a progress in geometrization of physics is possible
only on the way of {\it qualitative\/} modifications of the existing
space-time model, which requires a geometry qualitatively more general than
the Riemannian one. For example, a geometry in which Riemannian space
is tangential to the space of the new model.

The Finslerian 4D model could be a real candidate for this modification
\cite{Rund}. This geometry allows for a clear physical interpretation:
geometric properties of space may depend on the state of local classical
matter $(x^i,\dot x^k)$, not only on the coordinates (as in the
Riemannian model).

Although this model is not generally recognized in physics, it is used as a
background for many works on geometrization of physical phenomena,
including works devoted to some aspects of classical electrodynamics, e.g.,
\cite{Asan}. There are also many generalizations of the Finslerian model,
and the situation in this area, including the results, is similar to the
``Riemannian'' one.

Recently, the author managed to give a reason for such a situation with the
Finslerian model. The point is that the classical Finslerian geometry (as
well as its generalizations) is not relativistic \cite{Nos1}. It means that
this geometry cannot play the role of the geometry of a {\bf relativistic\/}
generalization of the modern Riemannian model.

For Clifford's hypothesis, this fact means a necessity of a) creating a
relativistic Finslerian geometry and b) proposing a new 4D space-time model
which could be described by this geometry. Here, the problem of geometrizing
electrodynamics should be considered as a test problem.

A new model of space-time\footnote{The model of embedded spaces (MES).}
and a simple metric version of the geometry were developed in \cite{Nos2,
Nos3}. The final formal effort in realization of this programme, derivation
of the field equations, is described in this paper.

It is clear that the required equations must be a set of local equations
for the field potentials of MES: an equation of Einstein type (for the
gravitational potential, the metric tensor $g_{ik}$) and an equation of
Maxwell type (for the electromagnetic potential, Cartan's torsion tensor
$C_{i,kl}$). (The equations must be generally covariant, satisfy the
correspondence principle and contain not higher than second-order
derivatives.) In spite of the fact that the potential of the developed
electromagnetic generalization is a third-order tensor, the prospective
source in the Maxwell-type equation will be the vector of electric current
density.

Besides, geometrization of fields requires a proper geometric understanding
of the method used to find these equations (the minimal action principle
in this case).

Thus, the aim of the work is derivation of MES field equations with, as far
as possible, a clear interpretation of both the method and results obtained.

\section{Lagrangian density\\ and action}

In our case, as in the construction of GR, it is absolutely natural to use
the curvature of MES geometry $\bar R$ to construct the field Lagrangian
density,
$$
    \bar R=g^{ik}\bar R_{ik},
\eqno(1)
$$
where the Ricci tensor is expressed in terms of the geometry connection as
\cite{Nos3}
$$
    \bar R_{ik}=\frac{\partial a^l_{\;ik}}{\partial x^l}-\frac{\partial a^l_{\;il}}{\partial x^k}
        +a^l_{\;ik}a^m_{\;lm}-a^l_{\;im}a^m_{\;kl}.
$$
Since the connection $a_{i,kl}=\Gamma_{i,kl}+\omega_{i,kl}$ is a sum of
the Christoffel and Lorentz terms,
$$
    \omega_{i,kl}=(F_{lm,ik}+F_{km,il}-F_{im,kl}) u^m,
$$
$$
    F_{ik,lm}=\partial C_{k,lm}/\partial x^i-\partial C_{i,lm}/\partial x^k,
\eqno(2)
$$
where
$$
    2C_{i,kl}=\partial g_{kl}/\partial u^i
\eqno(3)
$$
is Cartan's relativistic torsion tensor, the curvature (1) splits into
the Riemannian curvature $R$ (related to $\Gamma^i_{kl}$ in the
standard way) and the Lorentz $r$ parts
$$
    \bar R = R + r.
$$

In what follows, the Lorentz part of the curvature
$$
    r=2(\omega^{i,k}\;_{[k;i]}+\omega^{i,k}\;_{[k}\omega^l\,_{l]i})
$$
is conveniently expressed in terms of $F_{ik,lm}$, separating the
divergence term
$$
    r=4\left(F^{i[k,l]}\,_lu_i\right)_{;k}+2B_{ik}u^iu^k
$$
$$
    \equiv\frac{4}{\sqrt{-g}}\left(\sqrt{-g}F^{i[k,l]}\,_lu_i\right)_{,k}
            +2B_{ik}u^iu^k,
\eqno(4)
$$
where the symmetric tensor $B_{ik}$ is
$$
    B_{ik}=F^l_{\;(i,[m}\;^n\left(F_{lk),n]}\;^m-2F_{n]k),l}\;^m\right).
            %%(5)
$$
(The brackets $\mathstrut_{(.,.)}$ or $\mathstrut_{[.,.]}$ near indices
mean, as usual, symmetrization or anti-symmetri\-zation, respectively.)

It seems reasonable to use the scalar $\bar R_{ik}u^iu^k$ as a geometric
invariant which can also be used for building the desired Lagrange
density. However, this is not true: the Ricci tensor $\bar R_{ik}$ is a
function of the connection, and the anisotropy of the MES has already been
taken into account in the connection (2).

One of the basic assumptions of the MES is the concept of a congruence of
curves (trajectories of the initial matter congruence), where the
``initial'' (or ``bare'') matter means matter without contributions of its
own fields to inertia. To be more specific, let us assume that this matter
is distributed, with the densities of ``bare'' inertial mass and ``bare''
electric charge $\mu_0$ and $\rho_0$, respectively.  Moreover, let the
charge distribution be proportional to the mass distribution,
$$
        k=(\rho_0/\mu_0)^2,
\eqno(6)
$$
where $k$ is the gravitational constant.

At first sight, this assumption, unifying so strongly the initial matter,
has no reasonable grounds.  However, in what follows it will be
demonstrated that the field hypothesis guarantees renormalization of
$\rho_0$ and $\mu_0$ to values characteristic of dressed matter (moreover,
the assumption (6) allows the existence of neutral matter). Besides, the
following argument can be adduced: there exists the Eulerian description of
continuous matter, which is equivalent to the Lagrangian description. In
the framework of this description, the velocity of matter $u^i$ is treated
as a local field $u^i(x^k)$. Therefore, formally, the MES geometry may be
treated as a partial anisotropic case of Riemannian geometry.

Then, for the quantity $\rho_0/(\mu_0 c^2)$, in accordance with the
general principle of relativity, legitimate are only such values which
are combinations of the world constants (up to some numerical factor). The
quantity $(\rho_0/\mu_0)^2$ and the gravitational constant $k$ are
equidimensional, so that the simplest relation is (6).

Then the combination $-\bar R/2\varkappa$ (as in GR), with
$$
    \varkappa=8\pi kc^{-4},
\eqno(7)
$$
may be interpreted as the field Lagrangian 4-density (in any case, it
contains isotropic terms quadratic with respect to $\Gamma^i_{kl}$ for the
gravitational field).

Naturally, this interpretation can also be extended to the Lorentz term
$r$ of the curvature, especially to the terms quadratic with respect to
$F_{ik,lm}$. The fact that $F_{ik,lm}$ is included in $r$ only as a
contraction with $u^k$ can be treated as a result of MES anisotropy.

The integral action of the physical system also contains the free initial
matter term. Since this matter moves along geode\-sics of space, its
Lagrangian density must be chosen as
$$
    \Lambda_0=-\mu_0c^2ds/\sqrt{g_{00}}dx^0=-ci_{(0)}u,
\eqno(8)
$$
where
$$
    i^i_{(0)}=\mu_0cdx^i/\sqrt{g_{00}}dx^0
\eqno(9)
$$
is the current density of inertial mass of the initial matter.

Thus the sought-for action must include the following terms:
$$
    S\sim-c^{-1}\int_{\Omega}(\Lambda_0+\bar R/2\varkappa)
        \sqrt{-g}\, d\Omega,
\eqno(10)
$$
where $d\Omega$ is the 4-volume element.

Further, to formulate the variation problem, we need to define its
independent variables. At first sight, these should be the metric tensor
$g^{ik}$ (the gravitational field) and the tensor $C_{i,kl}$ (the
electromagnetic field). However, this supposition is wrong since they
{\it are not\/} independent quantities (see the definition (3)).
Therefore, the following geometric approach to the problem is valid.

As a curvature criterion at some point of Riemannian space, we may use a
scalar quantity, the interval $ds = \sqrt{g_{ik}(x^l)dx^idx^k}$, which is
the distance between this point and an infinitely close point (with
the coordinates $(x^i+dx^i)$). The interval is the length of a segment of
some curve passing through these points, and the segment itself is
situated along the unit tangential vector $u^i=dx^i/ds$ of the curve at
the point $(x^i)$. The Hilbert variation with respect to the metric
$g_{ik}$ (with respect to the squared linear point density of the
Riemannian space) means variation of this segment length for fixed
projections $dx^i$.

The case of MES space is more general: because of its anisotropy, it is
necessary to take into account the orientation of this segment relative to
a preferential direction at the point $(x^i)$ (relative to the curve of
the MES congruence which passes through this point). By virtue of the
isomorphism of MES curves $u^i\leftrightarrow u^i_{\rm mat}$
\cite{Nos2,Nos3}, the metric of MES space at the point $(x^i)$ can be
considered as $g^{ik}(x^l,u^m)$. Clearly, this conclusion is valid for both
the functional (10) and the particular case in which our curve belongs to
the congruence of MES curves (matter geodesics).

Thus such a generalization of the Hilbert variation to the case of MES
space is quite natural: it has two independent variations.

These are the ``old'' functional variation of (10) with respect to $g^{ik}$
and a ``new'' variation of (10) with respect to an {\it explicit\/}
dependence on $u_i$, because an implicit direction dependence is taken
into account by the first variation\footnote{This conclusion is confirmed by
the form of $r$, see (4) and (5): in the case of a linear dependence of
$g_{ik}$ on $u^l$, the field tensor $F_{ik,lm}$ depends only on the coordinates!}.

Such a choice of variables for the variation procedure makes it necessary
to consider the norms. It means that the present variation problem is
a problem with constraints imposed on the functional. Then the latter
must be written as
$$
    S=-c^{-1}\int_{\Omega}\Big[\Lambda_0 + (R + r)/(2\varkappa)
$$
$$
+ \lambda_1 g_{ik}g^{ik} + \lambda_2 u_iu^i\Big] \sqrt{-g}\,d\Omega,
\eqno(11)
$$
where $\lambda_1$ and $\lambda_2$ are Lagrange multipliers.

\section{Equation of Einstein\\ type}

As can be derived by varying (11) in $g^{ik}$ subject to the condition
that $\lambda_1$ and $\lambda_2$ are invariable constants in compliance
with the Lagrange method,
$$
0=\int_{\Omega}\Big[ \Big( R_{ik}-\frac{g_{ik}}{2}R-\varkappa t^{(m)}_{ik}+4u_{(i}B_{k)l}u^l
$$
$$
  -g_{ik} B_{lm}u^lu^m\Big) \delta g^{ik} + 2u^lu^m\delta B_{lm}\Big]\sqrt{-g}\,d\Omega,
$$
where
$$
    t^{(m)}_{ik} = t^{(0)}_{ik}-2\lambda_2u_iu_k+\left(2\lambda_1
        +\lambda_2\right)g_{ik},
\eqno(12)
$$
and $t^{(0)}_{ik} = -\Lambda_0u_iu_k$ is the energy-momen\-tum tensor (EMT)
of free initial matter.

Variation of the last term requires particular attention, but bearing in
mind that
$$
    \delta F_{ik,lm}\sim \frac{\partial^2 \delta g_{lm}}{\partial x^{[i}\partial u^{k]}}=0,
$$
because $\delta g_{lm}=0$, we can find the Einstein-type equation
in the form
$$
   R_{ik}-\frac{g_{ik}}{2}R=\varkappa\left(t^{(m)}_{ik}+t^{(em)}_{iklm}u^lu^m\right),
\eqno(13)
$$
where
$$
    t^{(em)}_{iklm}=-\frac4\varkappa\left(B_{iklm}+g_{(l(i}B_{k)m)}
        -\frac{g_{ik}}{4}B_{lm}\right)
\eqno(14)
$$
is the EMT of the electromagnetic field and the tensor $B_{iklm}$ is
$$
    B_{iklm}=F_{il,n}\,^{[p}F_{km,p}\,^{n]}+ 2\big( F_{nl,ip}F^{[n}\,_{m,k}\,^{p]}
$$
$$
    -F_{nl,ik}F^{[n}\,_{m,}\,^{p]}\,_p\big) + F_{(il,k)}\;^nF_{nm,p}\,^p
$$
$$
    + F^n\,_{l,n(i}F_{k)m,p}\,^p-2F_{(il,n}\,^pF^n\,_{m,pk)},
\eqno(15)
$$
so that both $B_{iklm}$ and $t^{(em)}_{iklm}$ are symmetric with
respect to indices inside the first and second pairs of indices.

\section{Equation of Maxwell type}

It is derived as an extremum condition of (11) in the directions
$\delta_{|u_i}S=0$.  The extremum is found from the explicit dependence of
$S$ on $u_i$ because the implicit dependence was already taken into account
in the Einstein-type equation. Thus we have
$$
    \int_{\Omega}\left[\delta r +2\varkappa\delta
       (\Lambda_0 + \lambda_2 u_i u^i) \right]\sqrt{-g}\,d\Omega=0.
$$
Using the expression (4) for $r$, taking into account the commutativity
of the operators $\partial/\partial u_i$ and $\partial/\partial x^k$,
$$
    \delta u_{i,k}=\delta u^l\frac{\partial^2u_i}{\partial u_l\partial x^k}=
        \delta u^l\frac{\partial^2u_i}{\partial x^k\partial u_l}=0,
$$
we obtain
$$
   \frac{2}{\sqrt{-g}}\left(\sqrt{-g}F^{i[k,l]}\,_l\right)_{,k}+2B^i\;_ku^k
$$
$$
        + \varkappa\left(\Lambda_0+2\lambda_2\right)u^i=0.
$$
But this equation has not a generally covariant form.

The requirement of general covariance can be satisfied if only this
equation is a set of equations:
$$
    F^{ik,l}\,_{l;k} + 2B^i\;_ku^k = -\varkappa(\Lambda_0+2\lambda_2) u^i,
$$
$$
    F^{il,k}\,_l = 0,
\eqno(16)
$$
where $F^{ik,l}\,_{l;k}\equiv\left(\sqrt{-g}F^{ik,l}\,_l\right)_{,k}/\sqrt{-g}$.

\section{Interpretation of the equations}

A physical meaning of the equations can be comprehended only after a
concrete definition of the MES ``vacuum'' concept. Partly, this definition
was discussed earlier \cite{Nos2}. Now, as a development of this notion,
based on the ideas of continuum matter used in this paper, it will be more
reasonable to define the MES ``vacuum'' as space areas of distributed
initial matter where $\rho_0\to 0$ and $\mu_0\to 0$. The ratio of these
quantities must satisfy (7).

Hence the vacuum of MES must have the properties of homogeneous and
extremely weakly charged inertial matter\footnote{A comparison of $q/m$ of some
charged elementary particle with $\sqrt{k}$ readily shows how weak is this
electric property of vacuum. E.g., for the proton we have $\sqrt{k}/(q_p/m_p) \simeq
10^{-21}$.},  moving with the velocity $u^i$. The gravitational constant, more
precisely, its algebraic square root with a certain (yet unknown) sign
$$
    \pm\sqrt{k} = \rho_0/\mu_0
\eqno(17)
$$
is the main {\it characteristic\/} of MES vacuum, like the Plank constant
for the density of space points \cite{Nos3}.

Such an approach allows us to treat the Universe as an area of space with
some fixed sign\footnote{Indirectly, this definiteness of the sign is confirmed by the
Universe asymmetry with respect to the content of matter and antimatter.} in (17).

Note that if we adopt this hypothesis, then, as a model of a charged
particle, we can choose a thermal vacuum fluctuation which has a long
lifetime due to its own fields preventing its decay. The fluctuation may
have the above $\rho_0$ and $\mu_0$ densities, but the condition (17) for
it holds true.

Thus in this model
$$
    \Lambda_0\neq 0,\quad  t^{(0)}_{ik}\neq 0,\quad   i^i_{(0)}\neq 0.
$$

Further we must understand how the initial matter is ``dressed'' with
fields.  In other words, since the role of the EMT of dressed matter is
played by $t^{(m)}_{ik}$ (12),
$$
    t^{(m)}_{ik}=-\left(\Lambda_0+2\lambda_2\right)u_iu_k
        +\left(2\lambda_1+\lambda_2\right)g_{ik},
\eqno(18)
$$
it is necessary to find a relationship of the Lagrange factors $\lambda_1$
and $\lambda_2$ with the field quantities. To do so, it is convenient to
use the Einstein-type equation in the following form:
$$
        t^{(m)}_{ik}=\left(R_{ik}-g_{ik}R/2\right)/\varkappa
                -t^{(em)}\,_{iklm}u^lu^m.
$$
Contractions with $g_{ik}$ and $u^iu^k$ lead to equations for the Lagrange
factors:
$$
    4\lambda_1+\lambda_2=\left(\Lambda_0-R/\varkappa
                -t^{(em)i}\,_{ikl}u^ku^l\right)/2,
$$
$$
    2\lambda_1-\lambda_2=\Lambda_0+\left(R_{ik}u^iu^k-R/2\right)/\varkappa
$$
$$
        -t^{(em)}\,_{iklm}u^iu^ku^lu^m,
$$
whence it follows
$$
    6\lambda_1 = 3\Lambda_0/2 - R/\varkappa + \big(R_{ik}/\varkappa
$$
$$
    - t^{(em)}_{iklm}u^lu^m-t^{(em)l}\,_{lik}/2 \big) u^iu^k,
$$
$$
      6\lambda_2=-3\Lambda_0+R/\varkappa- \big (4R_{ik}/\varkappa
$$
$$
      -4t^{(em)}_{iklm}u^lu^m+t^{(em)l}\,_{lik}\big )u^iu^k.
\eqno(19)
$$
The meaning of these formulae is clear: they describe geometrized matter.

Comparison of (19) with (18) for the EMT of dressed matter leads to a firm
conclusion that the geometrized EMT of matter is independent of the
Lagrange density of the initial (bare) matter $\Lambda_0$(!), but is
completely determined by the fields (gravitational and electromagnetic)
and by the matter velocity field (congruence of MES curves). (After
substitution of (19) into (18), $\Lambda_0$ vanishes from the coefficients
before $u^iu^k$ and $g_{ik}$.)

This result is also true for the Maxwell-type equation (16).

Consequently, the initial (``bare'') matter concept is just a redundant
though rather convenient hypothesis.

This conclusion should be interpreted as a direct proof of Clifford's
hypothesis for MES space: the MES field equations {\it do not include\/}
non-geometrical quantities (except the constant $k$).

Turning back, let us analyze the results of abandoning the initial
matter hypothesis.

First of all, it means that the first term in the action $S$ (11) of the
``matter + field'' physical system is unnecessary. It can be set to zero
or excluded from the Lagrange density, and the last two terms of the
action $S$ (with Lagrange factors) describe the matter terms:
$$
    6\lambda_1=-R/\varkappa + \big (R_{ik}/\varkappa
$$
$$
    - t^{(em)}_{iklm}u^lu^m-t^{(em)l}\,_{lik}/2\big) u^iu^k,
$$
$$
    6\lambda_2 = R/\varkappa - \big(4R_{ik}/\varkappa
$$
$$
    - 4t^{(em)}_{iklm}u^lu^m + t^{(em)l}\,_{lik}\big)u^iu^k,
\eqno(19')
$$
which means that to solve the set of field equations (13) and (16),
it is sufficient to know the vector field of matter velocities $u^i$
and the value of gravitational constant.

Secondly, the minimal action principle has an almost geometric meaning
(this meaning will be exactly geometric if it will be proved that the
gravitational constant $k$ has a geometrical origin).

Thirdly, the question of measurability of the scalar field $\Lambda_0$
(physically very nontrivial) is closed.

Fourthly, the concept of ``matter'' reduces to particle-like solutions of
the fields equations (more precisely, to the areas of these solutions
which have large curvature). In this case, the model of vacuum simply
suggests that there exist areas of space with minimal (zero as a limit)
curvature.

The only characteristic of the MES vacuum (in any case, for molar and
macroscopic scales) is the gravitational constant $k$.

And finally, the MES congruence of curves should be interpreted as world
lines of areas (points in the limit) with great curvature of particle-like
solutions.

Conclusion: the disavowal of the initial matter hypothesis has found such
a convincing geometric and physical justification that its further usage
would be a mistake.

The form of the matter EMT (18) shows that this matter can be treated as
a perfect fluid,
$$
    t^{(m)}_{ik} = (\varepsilon + p) u^iu^k - pg_{ik},
\eqno(20)
$$
where $\varepsilon$ and $p$ are the energy density and pressure, respectively.

So formally, by virtue of $\varepsilon$ and $p$ measurability, equation (13) can be
treated as a standard gravitational field equation with a source, which
includes two terms. The first term is the matter EMT (20) and the second
one is $t^{(em)}_{iklm}u^lu^m$.

Assuming that the quantities $\varepsilon$ and $p$ are known, their relationship
with the Lagrange factors are easily found by comparing (20) with (18) at
$\Lambda_0=0$:
$$
    2\lambda_2 = -(\varepsilon+p),\qquad  2\lambda_1 + \lambda_2 = -p.
$$

Let us substitute this result to the first Eq.(16) (at $\Lambda_0=0$):
$$
    F^{ik,l}\,_{l;k}+2B^i\;_ku^k=\varkappa (\varepsilon+p) u^i.
$$

In compliance with the vacuum model and Eq.(17), the tensor $F_{ik,lm}$
is related to the dimensional tensor $f_{ik,lm}$\footnote{Clearly, the role of $\alpha$
\cite{Nos2} is now played by $\alpha_0\equiv\pm c^{-2}\sqrt{k}$.} by
$$
    F_{ik,lm}=\pm c^{-2}\sqrt{k}f_{ik,lm},
\eqno(21)
$$
hence this equation can be rewritten as (see (7))
$$
    f^{ik,l}\,_{l;k}\pm 2c^{-2}\sqrt{k}\,b^i_ku^k
            =\pm 8\pi c^{-2}\sqrt{k} (\varepsilon + p) u^i,
\eqno(22)
$$
where $b_{ik}\equiv B_{ik}/kc^{-4}$ and
$$
   b_{ik}=f^l_{\;(i,[m}\;^n\left(f_{lk),n]}\;^m-2f_{n]k),l}\;^m\right).
\eqno(23)
$$

Now the analogy between Eq.(22) and the equation of Maxwell's
electrodynamics is obvious: the Maxwell field tensor $f^{ik}$ corresponds
to the contraction $f^{ik,l}\,_l$, and the electric current density $j^i$
is represented by
$$
    j^i = \mp 2c^{-1}\sqrt{k} (\varepsilon + p) u^i,
\eqno(24)
$$
from which directly follows the expression for the charge density
$$
       \rho=\mp2c^{-2}\sqrt{kg_{00}}(\varepsilon + p) u^{0},
\eqno(25)
$$
which can be named the field model of the electric charge. Obviously,
the equation of state of neutral matter is
$$
    \varepsilon+p = 0.
$$

It seems obvious that any change in the electric charge of a physical
system leads to a change of its energy and pressure because a) this is
related to variation of the matter density (the charge carriers are
particles, and so variation of the system charge is a change in the number
of its particles), i.e., variation of both $\varepsilon$ and $p$; b) charged
massless particles are unknown and c) a change in the charge leads to a
change in the system field, i.e., additional changes in the system energy
density and pressure. Hence the charge density, the energy density and
pressure should be locally interrelated. The only questionable point is
that this relation is qualitatively similar to (25). Note that (24) and (25)
are a classical (although relativistic) formulas, whereas the charge carriers
used in the experiment are quantum objects. Therefore we cannot state
that it is also valid in the quantum case. This will certainly require
experimental verification.

Here we must make a remark. At first sight, if a model of the electric
charge is a corollary of the field hypothesis, why, in this case, we
cannot treat the nonlinearity of the first equation (16) (or (22)) as an
additional term in the right-hand side, defining an electrical current
density?  After all, it is a contraction with velocity. The answer is
simple:  the current density, up to a {\it scalar\/} factor, must coincide
with the velocity of charged matter (as it is defined in electrodynamics).

The second equation of the set (16)
$$
    f^{il,k}\,_l = 0
$$
has no analogue in vectorial electrodynamics.

It can be demonstrated that the set (16) determines the tensor $F_{ik,lm}$
with completeness of Maxwell's electrodynamics.

First, for the antisymmetric pair of indices we have the first equation of
the system, which has the classical structure of the Maxwell equation (its
nonlinearity is inessential).

Second, for each fixed antisymmetric pair of indices, there are 10
components of the symmetric pair of tensor indices. For their
determination, this equation gives only one condition. That is, to
completely determine the tensor, one needs 9 more independent conditions,
which are given by the second equation (16). Really, the tensor
$F^{il,k}\,_l$ has no symmetries with respect to free indices. Therefore,
this equation gives 16 additional conditions. However, only 9 of them are
independent because the symmetric pair of indices of the tensor
$F_{ik,lm}$ are the indices of the metric tensor. The latter should always
satisfy 7 conditions for a choice of the reference frame (4 of them are
the conditions of choosing a spatial point and the other 3 are the
conditions of choosing the direction of motion at this point).

Finally, we must mention the identity
$$
    \partial F_{ik,mn}/\partial x^l+\partial F_{li,mn}/\partial x^k+
        \partial F_{kl,mn}/\partial x^i=0,
$$
which has a generally covariant form. This can be easily proved using the
orthogonality property of the potential $C_{i,kl}$ and the connection
$a_{i,kl}$ \cite{Nos2}.

\section{Discussion}

Construction of the field equations formally means that the test problem
of geometrization of classical electrodynamics has a solution in the
framework of the 4D MES. This solution is a ``rich'' generalization of
classic electrodynamics and leads to some serious conclusions on such
fundamental concepts as matter, electric charge, vacuum etc.
At the same time, the field hypothesis of matter comes to the fore, leaving
no other treatment of the latter. It is the author's strong belief that
this result is the only logically admissible result of physical treatment
of Clifford's idea.

Identification of the fields $F^{ik,l}\,_l$ with electromagnetic fields
will hardly cause any doubt. But identification of the residuary 9
additional fields of the tensor $F_{ik,lm}$, e.g., with classical
analogues of the Yang-Mills fields\footnote{In the classical case, $F_{ik,lm}$
has no non-Abelian terms because $C_{i,kl}$ is an unobservable quantity.
Therefore we speak about a ``classical analogue''. The quantum case of motion \cite{Nos3}
reduces to the particle ``scanning'' of its trajectory $\varepsilon\neq 0$.
Here $C_{i,kl}$ is already an observable quantity, and thus $F_{ik,lm}$ becomes
non-Abelian.  However, construction of a geometry for this case (as a generalization
of the one used) is still an unresolved problem.}, requires a separate investigation.

Of particular interest will be the results of MES investigations on
nonclassical scales. They will possibly allow us to unravel the puzzles of
the origin of the gravitational constant, the only physical constant of
the theory. Similar results will probably be obtained for other
fundamental constants (see, e.g., the considerations in \cite{Nos3}
concerning Planck's constant). In any case, Clifford's hypothesis suggests
a necessity of obtaining these answers.

Evidently, the predictions of the developed theory need experimental
verification. An electromagnetic ``redshift'' experiment was discussed
earlier, see, e.g., \cite{Nos2}. Now there appeared at least two new
corollaries: a non-field contribution to the mass of matter particles
is equal to zero, and the dependence of the charge density on the system
energy density and pressure. These two predictions have a principal
meaning for both the theory and Clifford's hypothesis.

\end{document}